\documentclass[12pt,preprint]{aastex}

\shorttitle{Non-Gaussianity with the SMHW}
\shortauthors{Vielva et al.}

\begin{document}

\title{Detection of non-Gaussianity in the WMAP 1-year data \\ 
using spherical wavelets}

\author{P. Vielva$^{1}$\footnotetext{e-mail: vielva@ifca.unican.es},
E. Mart{\'\i}nez-Gonz{\'a}lez$^{1}$,
R. B. Barreiro$^{1}$, J. L. Sanz$^{1}$ and L. Cay{\'o}n$^{1,2}$}

\affil{(1) Instituto de F{\'\i}sica de Cantabria,  
             Fac. de Ciencias, Av. de los Castros s/n, \\
             39005-Santander, Spain \\
 (2) Department of Physics, Purdue University, 525 Northwestern Avenue,
             West Lafayette, IN 47907-2036, USA}

\begin{abstract}

A non-Gaussian detection in the WMAP 1-year data is reported. The
detection has  
been found in the combined Q-V-W map proposed by the WMAP team
(Komatsu et al.  
2003) after applying a wavelet technique based on the Spherical Mexican Hat 
Wavelet (SMHW). The skewness and the kurtosis of the SMHW coefficients are 
calculated at different scales (ranging from a few arcmins to tens of degrees). 
A non-Gaussian signal is detected at scales of the SMHW around $4^\circ$ (size 
in the sky of around $10^\circ$). 
The right tail probability of the detection is $\approx 0.4\%$.
In addition, a study of Gaussianity is performed in
each hemisphere. 
The northern hemisphere is compatible with Gaussianity, whereas the
southern one  
deviates from Gaussianity with a right tail probability of $\approx 
0.1\%$.
Systematics, foregrounds and uncertainties in the estimation of the
cosmological  
parameters are carefully studied in order to identify the possible source of 
non-Gaussianity.
The detected deviation from Gaussianity is not found to be caused by
systematic effects: 1) 
each one of the Q, V and W receivers shows the same non-Gaussianity
pattern, and  
2) several combinations of the different receivers at each frequency
band ---that highly 
reduce the CMB and the foreground emissions--- do not show this non-Gaussian 
pattern. 
Similarly, galactic foregrounds show a negligible contribution to the
non-Gaussian detection: non-Gaussianity is detected in all the WMAP maps 
(from 23 GHz to 94 GHz) and no frequency dependence is observed.
Moreover, the expected foreground contribution to the combined WMAP map was 
added to CMB Gaussian simulations showing a behaviour compatible with the Gaussian 
model. 
Influence of uncertainties in the CMB power spectrum estimation are
also quantified.
Hence, possible intrinsic temperature fluctuations (like secondary anisotropies 
and primordial features) can not be rejected as the source of this non-Gaussian 
detection.
We remark that our result implies not only asymmetries north/south ---like other 
previous WMAP analyses--- but also a direct non-Gaussian detection.

\end{abstract}


\section{Introduction}

The study of the Gaussianity of the Cosmic Microwave Background (CMB) is one of 
the most powerful mechanisms for understanding the nature of the primordial 
density fluctuations: by estimating the probability distribution from the 
measured temperature fluctuations of the CMB, several models for the primordial 
density fluctuations can be rejected or accepted at a certain significance level.
For instance, standard inflationary models predict that the temperature 
fluctuations of the CMB correspond to a Gaussian, homogeneous and isotropic 
random field; whereas non-standard inflation (Linde \& Mukhanov 1997, Peebles 
1997, Benardeau \& Uzan 2002 and Acquaviva et al. 2002) and topological defects 
models (Turok \& Spergel 1990 and Durrer 1999) usually predict non-Gaussian 
random fields. Moreover, Gaussianity analyses can also be used 
to study the effect of
secondary anisotropies produced by the reionization of the 
universe (Ostriker \& Vishniac 1986 and Aghanim et al. 1996), the Rees-Sciama 
effect due to the non-linear evolution of the large scale structure (Rees \& 
Sciama 1968 and Mart{\'\i}nez--Gonz{\'a}lez \& Sanz 1990), gravitational lensing  
(Mart{\'\i}nez--Gonz{\'a}lez et al. 1997, Hu 2000 and Goldberg \&
Spergel 1999), etc. One of the handicaps within the detection of non-Gaussian 
signatures is related to the contamination of the CMB signal due to foregrounds 
(like the Galactic emissions and the compact sources).
These components produce \emph{spurious} non-Gaussian signals that can 
contaminate any intrinsic signature.

In order to uncover the nature of the primordial density 
fluctuations (or to conclude that the anisotropies are secondary), it  is 
essential to have high-resolution, low-noise and large-coverage CMB data.
Recently, the Wilkinson Microwave Anisotropy Probe (WMAP) NASA satellite has 
reported results from the 1-year all-sky data (Bennet et al. 2003a).
The WMAP science team (Komatsu et al. 2003) found the data to be consistent with 
Gaussianity. Two statistics were used: a measure of the phase correlations of 
temperature fluctuations (taking into account several combinations of the 
bispectrum) and the Minkowski functionals; moreover, improved limits for the 
non-linear coupling parameter were established.
Other groups have presented different analyses of Gaussianity on WMAP data:
Colley \& Gott (2003) have done an independent genus topology (one of the three 
Minkowski functionals) study, by performing a stereographic projection of the 
WMAP data, finding also Gaussian compatibility.
Chiang et al. (2003) presented a Gaussianity study (using a statistic based on 
the random-phase hypothesis) of the CMB map obtained from the WMAP data by 
Tegmark et al. (2003).
They found a significant non-Gaussian contribution at high-order multipoles. The 
authors conclude that this non-Gaussianity is due to unresolved foreground 
contamination. Gazta{\~n}aga \& Wagg (2003) have performed a 3-point angular 
correlation function analysis, finding good agreement with the Gaussian 
hypothesis. Higher-order moments 
(up to six) of the angular correlation function were calculated by Gazta{\~n}aga 
et al. (2003) which show the same Gaussian agreement. Another genus-statistic 
study has been done by Park (2003) using the foreground cleaned frequency maps 
of the WMAP data; a stereographic projection is also used finding a non-Gaussian 
detection by comparing the difference of the negative and positive genus ---with 
respect to the Gaussian prediction--- at different threshold levels. Several 
sources are considered in order to explain this detection, but higher 
signal-to-noise ratio data is needed to distinguish among them.
Eriksen et al. (2003) have computed 2- and 3-point correlations finding 
north/south asymmetries in the WMAP data. These asymmetries appear at large and 
intermediate scales and are associated with a lack of structure in the northern 
hemisphere, while the southern one is compatible with the Gaussian model 
(additional asymmetries east/west are also found)\footnote{During the refereeing 
process new works have appeared in astro-ph like Eriksen et al. 2004.}.

In this work we present a wavelet analysis of the Gaussianity of the WMAP 1-year 
all-sky map performed in wavelet space. Wavelets have been extensively used in 
the analysis of the CMB data, not only concerning Gaussianity studies, but also 
in the component separation field (Tenorio et al. 1999, Cay{\'o}n et al. 2001, 
Vielva et al. 2001a, Vielva et al. 2001b, Vielva et al. 2003 and Stolyarov et 
al. 2003) and in denoising techniques (Sanz et al. 1999a,b).
Regarding the application of wavelets to Gaussianity studies, Hobson et al. 
(1999) have shown that the wavelet coefficients provide a significantly better 
detection of non-Gaussian features due to cosmic strings than the Minkowski 
functionals.
Similarly, Aghanim et al. (2003) have shown that wavelets are more efficient to 
detect certain non-Gaussian features than other typical estimators like those based
in 
Fourier analysis (bispectrum and trispectrum).
Even more, due to the special nature of wavelets, the non-Gaussian sources can 
not only be detected but also identified in the maps, since the spatial 
information is kept in the wavelet space. Moreover, as it was shown in Barreiro 
\& Hobson (2001) and in Mart{\'\i}nez--Gonz{\'a}lez et al. (2002), the optimal 
combination of the information given by the wavelet coefficients at different 
scales, highly increases the power of wavelets to detect weak non-Gaussian 
signals.

The first application of wavelet techniques to detect non-Gaussian signatures 
was done by Pando et al. (1998), where the 2D flat Daubechies-4 wavelet was used 
to analyse the COBE-DMR data (see Mukherjee et al. 2000 for a critical review of 
this work). Lately, several works have used wavelets to study the Gaussianity of 
the COBE-DMR maps in the HEALPix scheme (G{\'o}rski et al. 1999): 
Barreiro et al. (2000) have used the Spherical Haar Wavelet (SHW), whereas 
Cay{\'o}n et al. (2001, 2003) have used the Spherical Mexican Hat Wavelet (SMHW). 
A critical comparison between the performances of both spherical wavelets for 
detecting non-Gaussian features (skewness and kurtosis) was done by 
Mart{\'\i}nez--Gonz{\'a}lez et al. (2002).

The aim of this paper is to apply the technique developed in Cay{\'o}n et al. 
(2001) and Mart{\'\i}nez--Gonz{\'a}lez et al. (2002) to the 1-year all-sky WMAP 
data in order to detect non-Gaussian features.
It is organized as follows. In Section~\ref{sec:simudata} the process to 
generate the simulations and to reduce the data is explained. In 
Section~\ref{sec:tool} the SMHW is briefly described and the different 
statistics are defined.
The SMHW analysis is presented in Section~\ref{sec:SMHW}, and the possible 
sources of the detected non-Gaussianity are discussed in 
Section~\ref{sec:sources}. Finally the conclusions are given in 
Section~\ref{sec:final}.
\label{intro}

\section{The WMAP data reduction and WMAP-like simulations}
\label{sec:simudata}
\begin{figure}
\begin{center}

   \includegraphics[width=16cm]{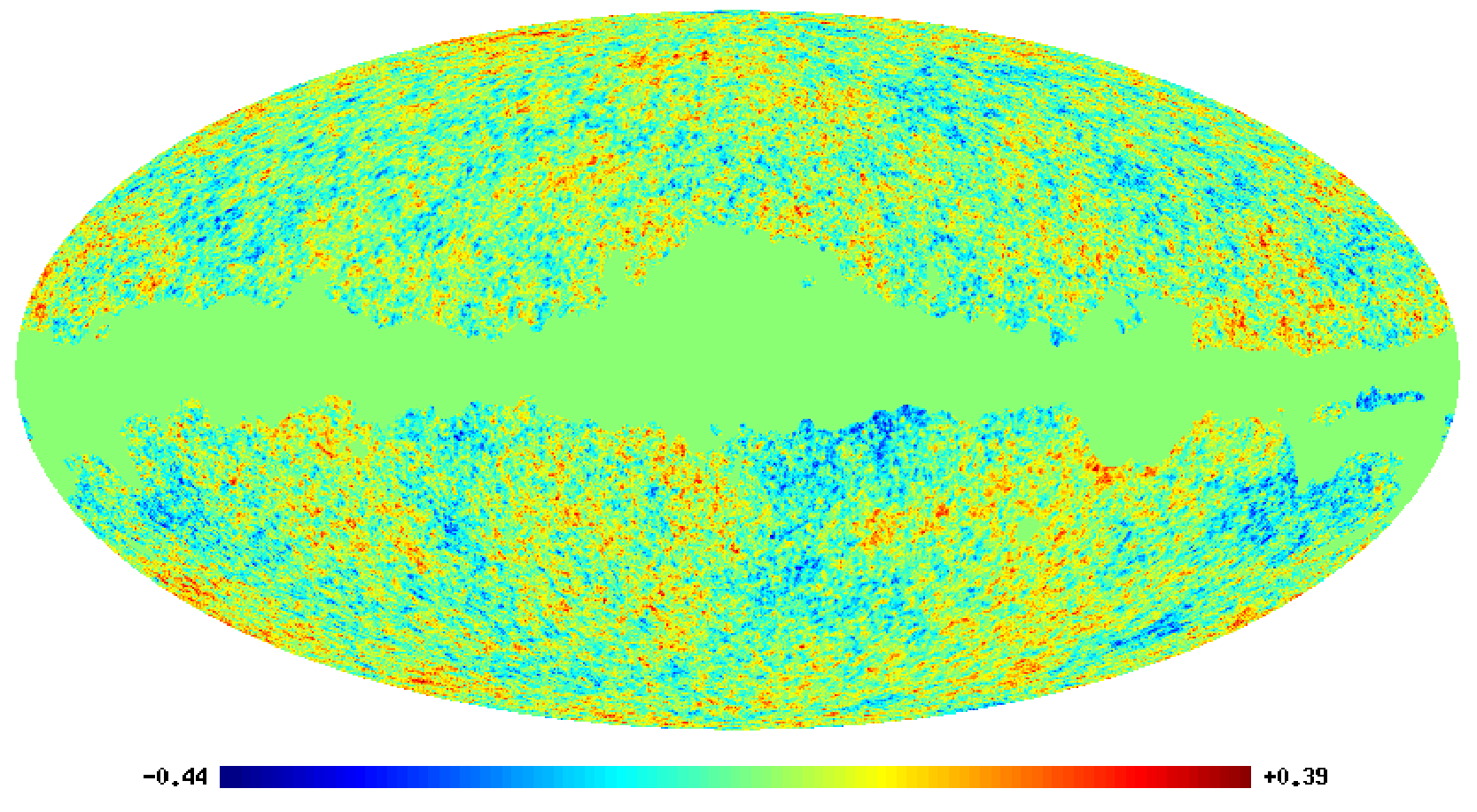}
 
\end{center}
\caption{\label{fig:WMAPDATA} Analysed WMAP map at $N_{side} = 256$
(in $mK$ units). It is a combination of 
all the receivers in Q-Band, V-Band and W-Band
after the foreground correction described in Bennett et
al. (2003b). The \emph{Kp0} mask has been  
applied and the residual monopole and dipole
have been subtracted.}
\end{figure}

The NASA Wilkinson Microwave Anisotropy Probe (WMAP) satellite was launched in 
the summer of 2001 and in February 2003 the 1-year results were presented. The 
WMAP radiometers observe at 5 frequency bands: K-band (22.8 GHz, 1 receiver), 
Ka-Band (33.0 GHz, 1 receiver), Q-Band (40.7 GHz, 2 receivers), V-Band (60.8 
GHz, 2 receivers) and W-Band (93.5 GHz, 4 receivers).
All the papers, data and products generated by the WMAP team can be found in the 
\emph{Legacy Archive for Microwave Background Data Analysis} (LAMBDA) web page
\footnote{http://cmbdata.gsfc.nasa.gov/}.
The noise and beam properties can also be found in the LAMBDA web page. 
The WMAP maps are presented in the Hierarchical, 
Equal Area and iso-Latitude Pixelization (HEALPix, G{\'o}rski et
al. 1999) at the  
$N_{side} = 512$ resolution parameter. 
The number of pixels is given by $12{N^2_{side}}$.
In order to perform the Gaussianity study, the WMAP team (Komatsu et
al. 2003)  
suggests combining all the maps produced by the receivers where the
CMB is the  
dominant signal (Q-Band, V-Band and W-Band).
We have followed the recommendation given by the WMAP team in the
choice of the combined map used in the analysis.
We would like to point out that none of the previous Gaussianity
studies of the WMAP data have used this combined map
(except the analysis performed by the WMAP team itself).
A very simple pipeline is given in Bennett et al. (2003b). At a given
position in the sky (${\mathbf{x}}$), the temperature is given by:
\begin{equation}
\label{eq:combination}
T({\mathbf{x}}) = \sum_{j = 3}^{10} 
{T_j}({\mathbf{x}})~{w_j}({\mathbf{x}}),
\end{equation}
where the indices $j = 3, 4$ refer to the Q-Band receivers, $j = 5, 6$
to the ones of the V-Band and, finally, $j = 7, 8, 9, 10$ correspond
to the receivers of the W-Band (the indices $j = 1, 2$ are used for 
the K and Ka receivers, respectively).
The noise weight ${w_j}({\mathbf{x}})$ is defined by:
\begin{eqnarray}
\label{eq:noise}
w_j({\mathbf{x}}) = \frac{\bar{w}_j({\mathbf{x}})}{\sum_{j =
3}^{10}{\bar{w}_j}({\mathbf{x}})},~~ &
\bar{w}_j({\mathbf{x}}) = \frac{{N_j}({\mathbf{x}})}{{{\sigma_0}_j}^{2}}
\end{eqnarray}
where ${{\sigma_0}_j}$ is the noise dispersion per observation and 
${N_j}({\mathbf{x}})$ is the number of observations made by the
receiver $j$ at position ${\mathbf{x}}$ (see Bennett et al. 2003a).
Equation~(\ref{eq:combination}) provides a single map where the
signal-to-noise  
has been increased. Although the CMB dominated frequencies have been chosen, we 
still have significant contribution due to Galactic foregrounds (thermal dust, 
free-free and synchrotron) as well as extragalactic point sources (negligible 
Sunyaev-Zel'dovich contribution due to galaxy clusters is expected, see Bennett 
et al. 2003b). In order to avoid the Galactic emissions, the WMAP team performed 
a foreground template fit described in Section 6 of Bennett et al. 2003b. The 94 
GHz dust map of Finkbeiner et al. (1999) is used as the thermal dust template; 
the $H_{\alpha}$ map of Finkbeiner (2003) corrected for extinction through the 
$E_{B-V}$ map of Schlegel et al. (1998) is used as the free-free template; 
finally, the synchrotron template is the 408 MHz Haslam et al. (1982) map.
Hence, Equation~(\ref{eq:combination}) is modified by:
\begin{equation}
\label{eq:combination2}
\hat{T}({\mathbf{x}}) = \sum_{j = 3}^{10} 
\hat{T_j}({\mathbf{x}})~{w_j}({\mathbf{x}}),
\end{equation}
where $\hat{T_j}({\mathbf{x}})$ is the temperature at position ${\mathbf{x}}$ 
for the receiver $j$ after foreground correction. The parameters for the 
best-fit of the foreground templates are given in Bennett et al. (2003b). 
Even more, the foreground cleaned frequency maps can be found in the LAMBDA 
site.
The map is then degraded to resolution $N_{side} = 256$ ---since the very 
small scales are dominated by noise--- and a mask is applied to avoid the 
contamination due to the strong emission at the Galactic plane and the 
contribution due to known radio point sources. This mask, that can
also be found in the LAMBDA site, is called
\emph{Kp0} and keeps 76.8\% of the sky.
Finally, the residual monopole and dipole outside the mask are removed.
The final map is shown in Figure~\ref{fig:WMAPDATA}.

In order to study the Gaussianity of this map, we have produced 10000
Gaussian  
simulations. Using CMBFAST (Seljak \& Zaldarriaga 1996), we have
calculated the  
$C_\ell$ given by the cosmological parameters estimated by the WMAP team (Table 1 of 
Spergel et al. 2003).
Random Gaussian $a_{\ell m}$ of CMB realizations have been generated
and convolved at  
each one of the WMAP receivers with the adequate beams.
After the transformation from harmonic to real space, uncorrelated Gaussian 
noise realizations have been added following the number of observations per 
pixel (${N_j}({\mathbf{x}})$) and the noise dispersion per observation 
(${{\sigma_0}_j}$).
We have combined all the maps following Equation~(\ref{eq:combination2}). 
Finally, the 10000 simulations have been degraded to $N_{side} = 256$, the
\emph{Kp0}  
mask has been applied, and the residual monopole and dipoles have been
fitted and subtracted for each simulation independently. 
\section{The tool and the non-Gaussian estimators}

\label{sec:tool}
Wavelet techniques have been shown to be very powerful for detecting
non-Gaussianity in CMB data (Hobson et al. 1999, Aghanim et al. 2003).
Due to the special nature of wavelets, a multi-scale study can be performed
to amplify the signature of the 
non-Gaussian features dominating at a given scale.
Moreover, the SMHW is ideal for the enhancement of non-Gaussian signatures with 
spherical symmetry.
The SMHW has been already applied for non-Gaussian studies to the
COBE-DMR data (Cay{\'o}n et al. 2001, 2003) and to \emph{Planck} simulations 
(Mart{\'\i}nez-Gonz{\'a}lez et al. 2002).
The SMHW can be obtained from the Euclidean Mexican Hat Wavelet (MHW) following 
the stereographic projection suggested by Antoine \& Vanderheynst
(1998). This projection ensures that the wavelet properties are kept
and that the MHW is recovered in the small angle limit (see
Mart{\'\i}nez-Gonz{\'a}lez et al. 2002  
for a graphical explanation of this extension).
The SMHW satisfies the \emph{compensation}, \emph{admissibility} and
\emph{normalization} properties that define a wavelet and is given by:
\begin{equation}
\label{eqSMHW}
   \Psi_S(y,R) = 
\frac{1}{\sqrt{2\pi}N(R)}{\Big[1+{\big(\frac{y}{2}\big)}^2\Big]}^2
  \Big[2 - {\big(\frac{y}{R}\big)}^2\Big]e^{-{y}^2/2R^2},
\end{equation}
where $R$ is the scale and $N(R)$ is a normalization constant:
\begin{equation}
\label{N}
      N(R)\equiv R{\Big(1 + \frac{R^2}{2} + \frac{R^4}{4}\Big)}^{1/2}.
\end{equation}
The distance on the tangent plane is given by $y$ that is related to the polar 
angle ($\theta$) through:
\begin{equation}
\label{y}
      y\equiv 2\tan \frac{\theta}{2}.
\end{equation}

At a given scale ($R$), several statistics can be defined. In this work we have 
used two simple non-Gaussian estimators: skewness ($S(R)$) and kurtosis 
($K(R)$):
\begin{eqnarray}
S(R) & = & \frac{1}{N_R} \sum_{i = 1}^{N_R} {w_{i}(R)}^3  {\Big /} {\sigma(R)}^3  
\\
K(R) & = & \frac{1}{N_R} \sum_{i = 1}^{N_R} {w_{i}(R)}^4  {\Big /} {\sigma(R)}^4 
- 3, 
\end{eqnarray}
where $N_R$ is the number of coefficients at scale R and
$\sigma(R)$ is the dispersion of the wavelet coefficients at the scale R 
($w_{i}(R)$):
\begin{eqnarray}
{\sigma^2(R)} & = & \frac{1}{N_R} \sum_{i = 1}^{N_R} {w_{i}(R)}^2 .
\end{eqnarray}%
In the previous $S(R)$, $K(R)$ and $\sigma(R)$ definitions, it is assumed that 
the $N_R$ wavelet coefficients ($w_{i}(R)$) have zero mean at each scale $R$.

\section{The Spherical Wavelet analysis}
\label{sec:SMHW}

In order to test the Gaussianity of the WMAP 1-year data, we have applied the 
following analysis using the SMHW.

We have performed 10000 simulations following the pipeline proposed by 
Bennett et al. (2003b) and already indicated in Section~2. Each one of these 
simulations have been convolved with the SMHW at different scales ($R_1 = 13.7$, 
$R_2 = 25$, $R_3 = 50$, $R_4 = 75$, $R_5 = 100$, $R_6 = 150$, $R_7 = 
200$, $R_8 = 250$, $R_9 = 300$, $R_{10} = 400$, $R_{11} = 500$, 
$R_{12} = 600$, $R_{13} = 750$, $R_{14} = 900$ and $R_{15} = 1050$ 
arcmin).
Acceptance intervals at certain significance levels $\alpha$ 
(e.g., 32\%,
5\% and 1\%) have been established at each scale based on these simulations. These acceptance intervals are defined as the intervals which
contain a probability $1-\alpha$ and the  
remaining probability is the same above and below the interval, i.e. 
$\alpha/2$ at each side.   
The acceptance intervals have been determined by studying
the distribution of the 
skewness and the kurtosis at each scale independently and calculating the
corresponding percentiles.
We have checked that the number of simulations performed (10000) is
enough to establish those acceptance intervals with good precision. 
Finally, the same analysis has been applied to the WMAP map
($\hat{T}({\mathbf{x}})$) plotted in Figure~\ref{fig:WMAPDATA}.

\begin{figure}
\begin{center}

   \includegraphics[width=16cm]{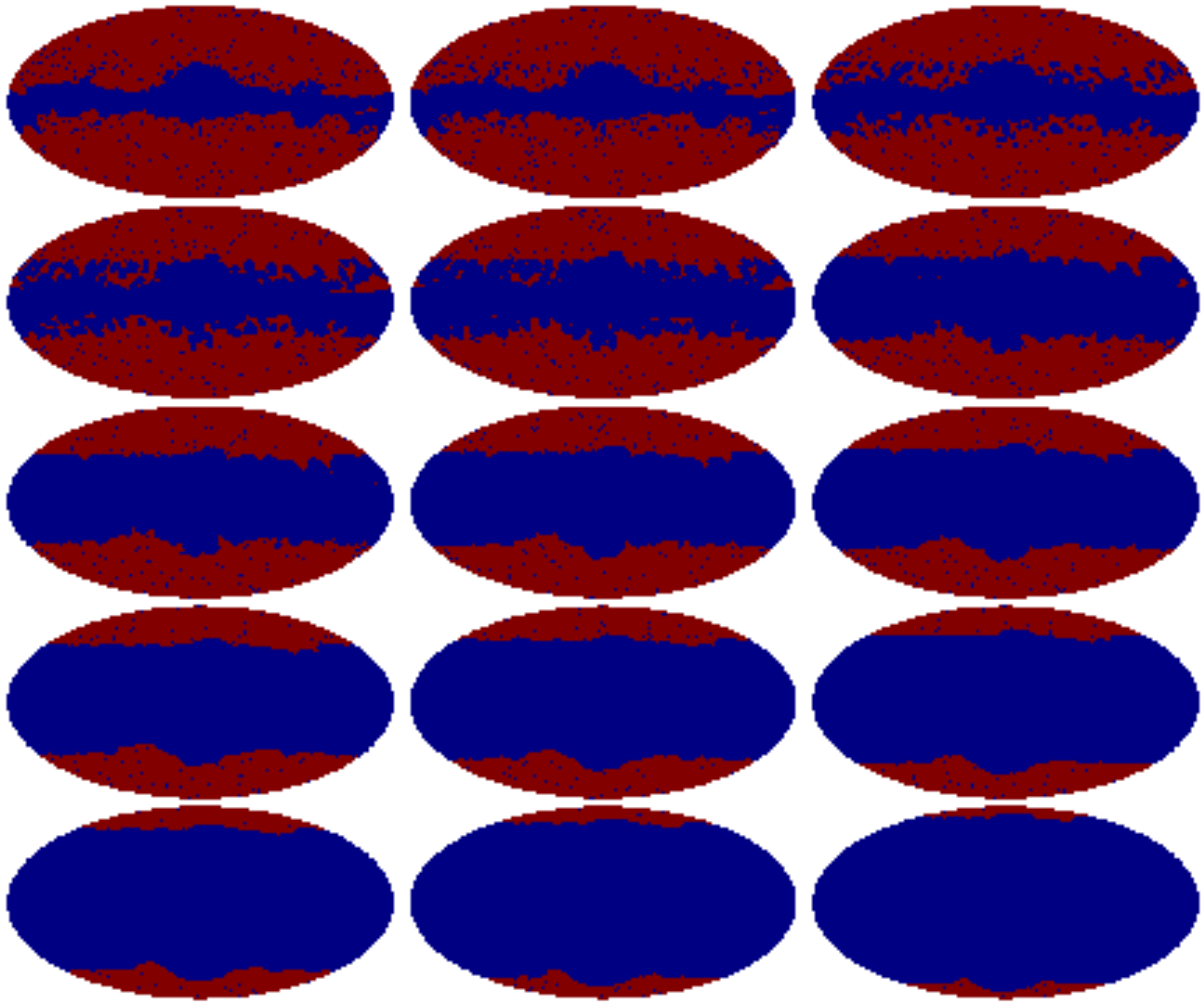}
 
\end{center}
\caption{\label{fig:SET1.MASKS}Set of \emph{exclusion masks} ($M(R)$)
used in our analysis. This set is defined in order
to discard in the SMHW analysis those pixels with a strong \emph{kp0} mask
contamination. 
Since the number of
excluded pixels grows with the SMHW scale, there is one
\emph{exclusion mask} per scale $R$. 
The \emph{exclusion masks} (from left to right and top to
bottom) correspond to the following scales: $R_1 = 13.7$, 
$R_2 = 25$, $R_3 = 50$, $R_4 = 75$, $R_5 = 100$, $R_6 = 150$, $R_7 = 
200$, $R_8 = 250$, $R_9 = 300$, $R_{10} = 400$, $R_{11} = 500$, 
$R_{12} = 600$, $R_{13} = 750$, $R_{14} = 900$ and $R_{15} = 1050$ 
arcmin}
\end{figure}

Since we are convolving a map with a large Galactic mask plus known point 
sources (the so-called \emph{Kp0} mask) with the SMHW, we are
introducing in the  
wavelet analysis a large number of pixels ---those near the border of
the mask---  
largely affected by the zero value of the mask.
This effect is taken into account through the simulations, but,
obviously, it is  
introducing an undesirable loss of efficiency. 
In order to avoid this effect, we will not consider pixels whose SMHW 
coefficients have a strong contamination from the mask. 
In other words, we \emph{exclude} these pixels after the SMHW convolution.
Since we are convolving the combined WMAP map 
($\hat{T}({\mathbf{x}})$) with different scales, the number of pixels to be 
excluded is growing with the scale.
There could be several criteria to define this set of \emph{exclusion masks}. 
We have tested several definitions and all of them lead to the same results. 
The results here presented were obtained with an \emph{exclusion mask}
defined  
as follows: at a given scale ($R$), the \emph{exclusion mask} $M(R)$ is composed 
by the \emph{Kp0} mask plus those pixels that are closer than $2.5R$ to any one 
of the pixels of the \emph{Kp0} mask that are in the Galactic plane. In 
other words, at a given scale, the \emph{exclusion mask} is an enlargement 
(proportional to the scale) of the \emph{Kp0} mask from the Galactic plane (but 
note that the mask is not increased around the masked point sources).
The \emph{exclusion masks} corresponding to the 15 scales are plotted in 
Figure~\ref{fig:SET1.MASKS}.

\begin{figure}
\begin{center}

   \includegraphics[angle=270,width=16cm]{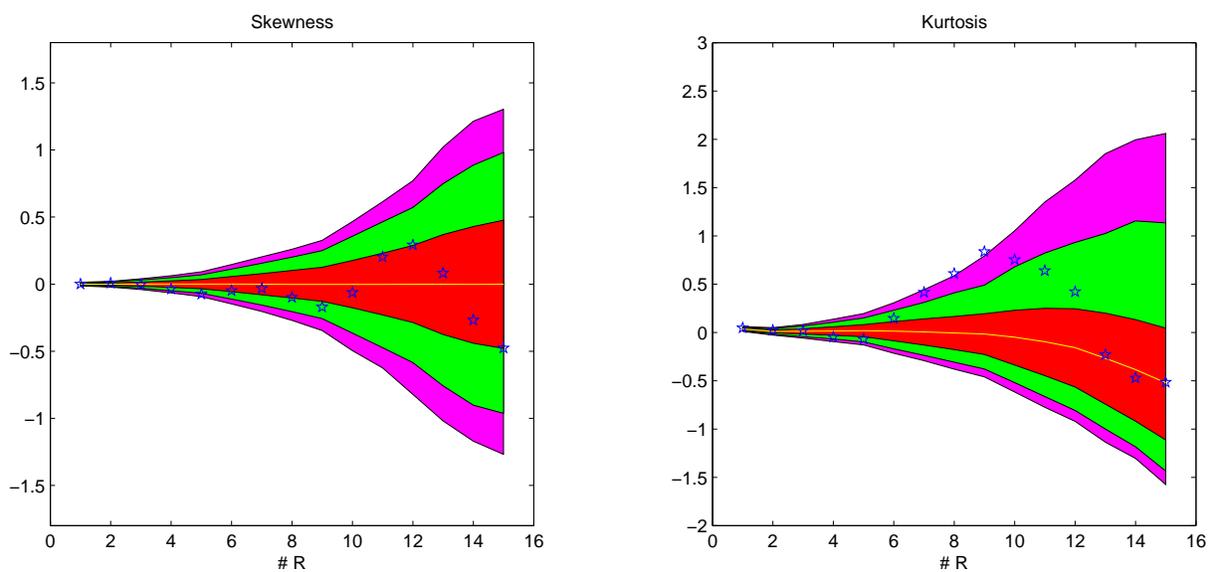}
 
\end{center}
\caption{\label{fig:graf_wmap_set1}The skewness ($S(R)$, left panel) and kurtosis
($K(R)$, right panel) 
values obtained from the application
of the SMHW analysis to the combined WMAP map ($\hat{T}({\mathbf{x}})$) are shown as
blue starts.
Acceptance intervals for the 32\% (red, inner), 5\% (green, middle)
and 1\% (magenta, outer) significance levels are also plotted, as well as the mean
value
given by the 10000 simulations performed in this work
(yellow solid line). Only those pixels allowed by the \emph{exclusion masks} set
($M(R)$) have been used in the analysis (see text). }
\end{figure}

Following this process we obtain the results presented in 
Figure~\ref{fig:graf_wmap_set1}: an excess of kurtosis is detected at two 
consecutive scales: $R_8 = 4.17^\circ $ and $R_9 = 5^\circ$. 
The value of the kurtosis at $R_8 = 4.17^\circ$ is a non-Gaussian detection
outside the acceptance interval at the $1\%$ significance level. 
Even more, only 40 of the 10000 simulations present a kurtosis value larger or
equal
than the one detected in the combined WMAP map at this scale. This 
corresponds to a right tail probability of $\approx 0.4\%$.
A similar result is found at scale $R_9 = 5^\circ$ for which 
the right tail probability is also $\approx 0.4\%$.

We want to remark that this value is obtained for very different definitions of 
the \emph{exclusion mask}.
For instance, we have tested an enlargement of the \emph{Kp0} mask from the 
Galactic plane scaling with the SMHW scale ($R$) as: $0.7R$, $0.8R$, $0.9R$, 
$1.0R$ and $1.75R$. Moreover, an enlargement of the whole \emph{Kp0} mask not 
only from the Galactic plane but also from all the masked point sources has 
been also checked, leading to similar results.

\begin{figure}
\begin{center}

   \includegraphics[width=12cm]{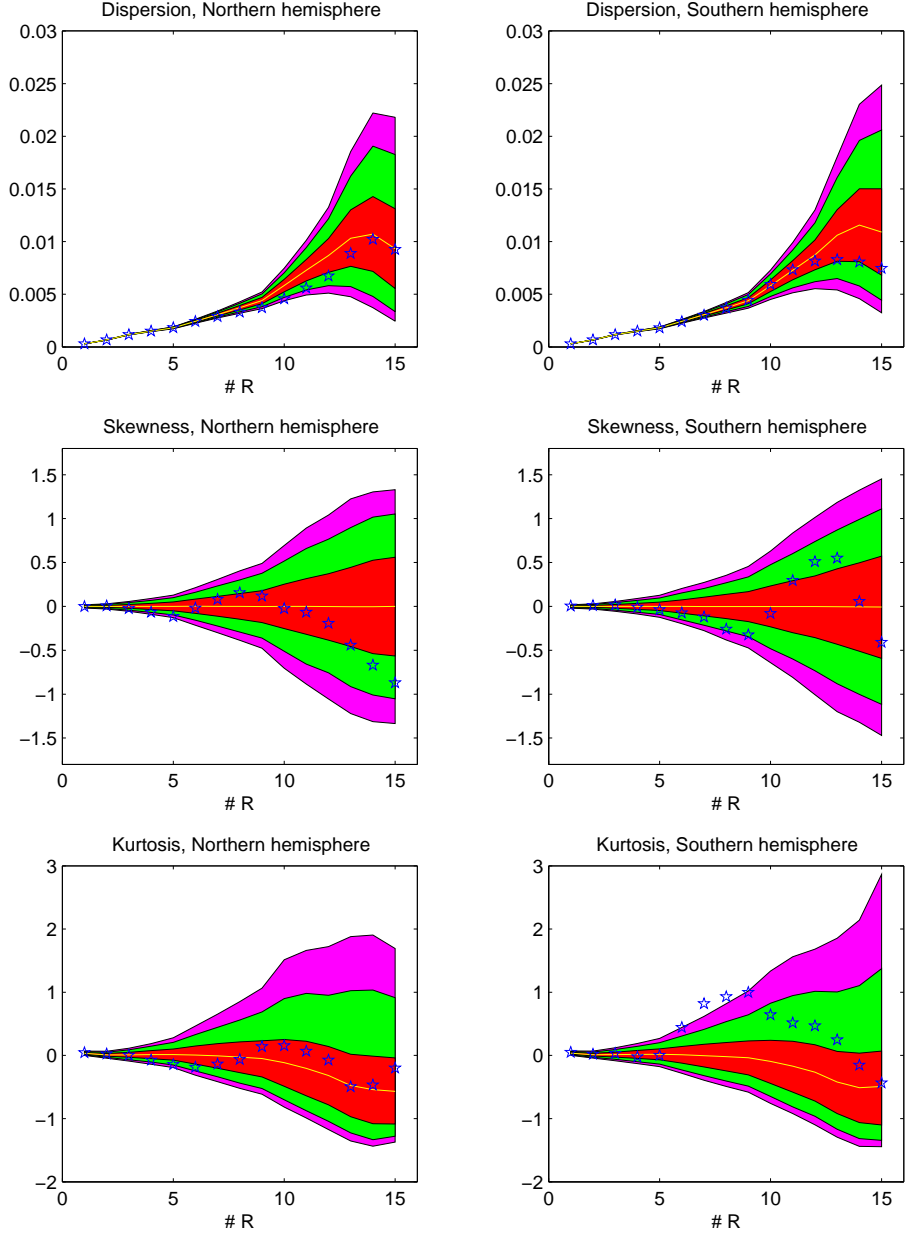}
 
\end{center}
\caption{\label{fig:graf_wmap_set1_NandS}The dispersion ($\sigma(R)$, top)
skewness ($S(R)$, middle) and kurtosis ($K(R)$, bottom) 
values obtained from the application
of the SMHW analysis to the combined WMAP map ($\hat{T}({\mathbf{x}})$) are shown
(blue starts).
The left column corresponds to the analysis performed only in the northern
hemisphere, whereas the analysis
of the southern one is in the right column.
Acceptance intervals for the  32\% (red, inner), 5\% (green, middle)
and 1\% (magenta, outer) significance levels are also plotted, as well as the mean
value
given by the 10000 simulations (yellow solid line). The \emph{exclusion masks} set
($M(R)$) have been used in the analysis (see text). }
\end{figure}

In addition we have performed the SMHW analysis in each hemisphere to
determine if the non-Gaussian detection is present in all the sky. 
Previous non-Gaussianity works on the WMAP data have reported some
asymmetries  
between the northern and the southern hemisphere (Park 2003, Eriksen et al. 
2003). As Figure~\ref{fig:graf_wmap_set1_NandS} clearly shows, the kurtosis is 
located in the southern hemisphere.
The non-Gaussian detection in the southern hemisphere occurs at the $R_7 = 
3.33^\circ$ and $R_8 = 4.17^\circ $ SMHW scales.
The value of the kurtosis at scale $R_7 = 3.33^\circ$ is 
above the acceptance interval at the $1\%$ significance level. 
Only 11 of the 10000 simulations present values for the 
kurtosis equal or larger than the one detected at this scale, which
corresponds to a right tail probability of $\approx 0.1\%$.
A similar result is found for the scale $R_8 = 4.17^\circ $ for which 
the right tail probability is $\approx 0.2\%$. 

By looking at the scalogram of the wavelet coefficients in the northern and 
southern hemispheres (dispersions at the different scales $R$ in 
Figure~\ref{fig:graf_wmap_set1_NandS}), we can see that the northern hemisphere 
shows less structure than the southern one, at scales between $4^\circ$ and  
$12^\circ$.
On the contrary, the northern hemisphere has more structure than the southern 
one at larger scales. However, the scalograms of both hemispheres are
within the acceptance intervals of the Gaussian model.

Let us remark that, as mentioned before, previous works have reported 
asymmetries between the northern and southern hemispheres. For instance, Park 
(2003) has found different north/south genus behaviour which implies a non-
Gaussian detection at the $99\%$. Eriksen et al. (2003) also found an
 asymmetric behaviour north/south regarding the structure level at large and 
intermediate angular scales: the northern hemisphere has a lack of structure 
whereas the southern one is compatible with the Gaussian model. 
However, our study reveals: 1) a direct measurement of non-Gaussian signatures 
related to the kurtosis of those structures with a typical size in the sky of 
around $10^\circ$, with a probability $\approx 0.4 \%$; 2) the southern 
hemisphere shows an excess of kurtosis with a probability $\approx 0.2 \%$ 
(around the previous scale) whereas the northern one is compatible with the 
Gaussian model.

Finally, we would like to point out that we have also performed a
wavelet analysis based  
on the Spherical Haar Wavelet (SHW) as it was proposed in Mart{\'\i}nez-
Gonz{\'a}lez et al. (2002). We did not find any non-Gaussian detection
using the  
SHW. This is not so surprising, since (as it was shown in Mart{\'\i}nez-
Gonz{\'a}lez et al., 2002) the SHW is less efficient than the SMHW to detect non-
Gaussian signals.

\section{Discussion: Sources of non-Gaussianity}
\label{sec:sources}

The detection of non-Gaussianity in the WMAP map ($\hat{T}({\mathbf{x}})$) was 
presented in the previous Section. The kurtosis of the wavelet coefficients at 
scales $R\approx 4^\circ$ was found to have a very low probability $\approx 0.4\%$. The 
non-Gaussianity detection is localized in the southern hemisphere.

The aim of this Section is to study, as far as we can, the source of this non-
Gaussianity. Let us summarise the major hypotheses that have been assumed:

\begin{enumerate}

\item The CMB is a homogeneous and isotropic multivariate Gaussian random field
on the sphere 
\item The WMAP data are free from systematics
\item The WMAP data outside the Kp0 mask are not contaminated by foregrounds
\item The uncertainties in the cosmological parameters have a negligible 
effect on the results

\end{enumerate}

Obviously, the most critical hypothesis is 1. If we find that
hypotheses 2 to 4  
have a negligible effect on the results, then we could conclude that
hypothesis 1 is not verified. In fact, hypothesis 1  
contains at least 2 hypotheses:

\begin{itemize}
\item The CMB is a homogeneous and isotropic random field on the sphere
\item The CMB is a multivariate Gaussian random field (or equivalently the 
$a_{lm}$ are Gaussian)
\end{itemize}

In practice it is very difficult to distinguish between these two hypotheses.
On the one hand, since we are using non-Gaussian estimators, 
we prefer to refer to a non-Gaussian detection which is produced by localised 
regions.
More precisely, we have found a non-Gaussian detection that seems to
be present  
in the southern hemisphere rather than in the northern one.
On the other hand, homogeneity and isotropy is a more fundamental
principle in cosmology than Gaussianity.

Systematic effects are a possible source of non-Gaussianity. In
Subsection~\ref{subsec:sytematics}, we will study if 
there are any systematic effects: are any of
the receivers producing the non-Gaussian feature?, or the noise?, or
the beam?.  
In Subsection~\ref{subsec:fore}  we present tests to study the possible 
influence of the
Galactic foregrounds on the detection. The influence of the accuracy of the 
estimated power spectrum of the WMAP data (or of the derived cosmological 
parameters) on our results is also studied (Subsection~\ref{subsec:param}). 
Finally, in Subsection~\ref{subsec:intrinsic}, we speculate about intrinsic 
fluctuations.

\subsection{Systematics}
\label{subsec:sytematics}

Our first test to check the influence of systematics in the non-Gaussian 
detection consists on looking for any \emph{rare} receiver. Instead of analysing 
the WMAP map ($\hat{T}({\mathbf{x}})$) obtained by the weighted average of all 
the Q-Band, V-Band and W-Band receivers, we perform the SMHW analysis in each 
receiver map ($\hat{T}_{Q_1}({\mathbf{x}})$, $\hat{T}_{Q_2}({\mathbf{x}})$,
$\hat{T}_{V_1}({\mathbf{x}})$, $\hat{T}_{V_2}({\mathbf{x}})$, 
$\hat{T}_{W_1}({\mathbf{x}})$, $\hat{T}_{W_2}({\mathbf{x}})$, 
$\hat{T}_{W_3}({\mathbf{x}})$ and $\hat{T}_{W_4}({\mathbf{x}})$) independently.
If the results are the same for these maps, then we can conclude that the 
detection of non-Gaussianity is not produced by any particular receiver. 

\begin{figure}
\begin{center}

   \includegraphics[width=6cm]{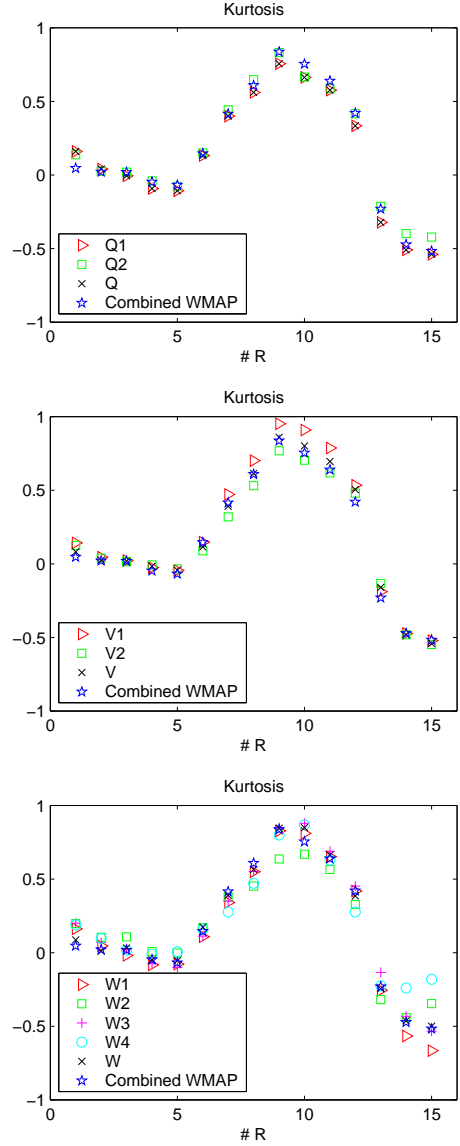}
 
\end{center}
\caption{\label{fig:graf_systematics1}Kurtosis values ($K(R)$) obtained from
the application of the analysis described in this work to 
each WMAP maps observed at channels: $Q_1$ and $Q_2$ (top panel), $V_1$ and $V_2$
(middle panel) and
$W_1$, $W_2$, $W_3$ and $W_4$ (bottom panel).
For a better comparison, the kurtosis obtained for the combined WMAP map has also
been included (blue starts)
in all the panels.}

\end{figure}

Results are presented in Figure~\ref{fig:graf_systematics1}. The $K(R)$ curve 
obtained for the WMAP map ($\hat{T}({\mathbf{x}})$) is plotted together with the 
curves obtained for each of the receivers ($K_{Q_1}(R)$, $K_{Q_2}(R)$, 
$K_{V_1}(R)$, $K_{V_2}(R)$, $K_{W_1}(R)$, $K_{W_2}(R)$, $K_{W_3}(R)$, 
$K_{W_4}(R)$).
We can see how the pattern for the $K(R)$ curve is almost perfectly
followed by the other kurtosis curves. 

\begin{figure}
\begin{center}

   \includegraphics[width=6cm]{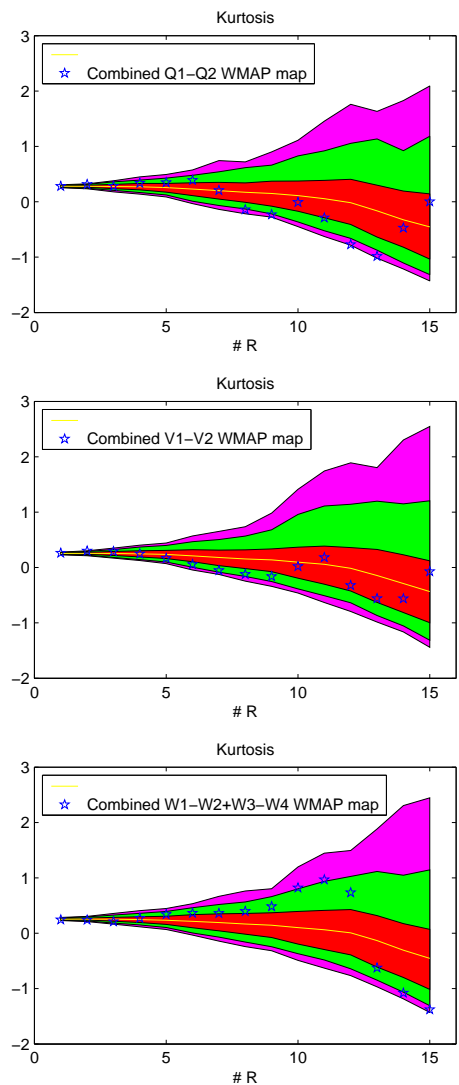}
 
\end{center}
\caption{\label{fig:graf_systematics2}Kurtosis values ($K(R)$) obtained from
the analysis of different receiver combinations (where the CMB and foreground
contributions
are negligible) are shown as stars: $Q_1-Q_2$ (top panel), $V_1-V_2$ (middle panel)
and
$W_1-W_2+W_3-W_4$ (bottom panel).
As in Figure~\ref{fig:graf_wmap_set1},
the acceptance intervals at the 32\% (red, inner), 5\% (green, middle) and
1\% (magenta, outer) significance levels for the same receiver combination 
and the mean value given by the 1000 simulations (yellow solid line) are
also shown. }
\end{figure}

We have also checked the influence of possible systematics related to
instrumental 
features (noise and beams). We have performed the SMHW analysis on
maps produced from  
the subtraction of receivers at the same frequency. These maps are
almost free  
from CMB and foreground contribution (remaining residuals can be present due to 
the slightly different resolutions between the two subtracted channels). In 
particular, we have analysed the maps obtained by subtracting the two
receivers  
in the Q-band ($Q_1 - Q_2$), the two receivers in V-Band ($V_1 - V_2$)
and the map obtained from $W_1 - W_2 + W_3 - W_4$. As seen in 
Figure~\ref{fig:graf_systematics2}, the patterns followed by the kurtosis in 
these three cases are compatible with simulations assuming 
a Gaussian CMB and performing the same operations as for the real data at each
frequency
(for each case, 1000 simulations were generated to establish the 
acceptance intervals at the 32\%, 5\% and 1\% significance levels). 
In addition, they are completely different from the $K(R)$ 
curve obtained for the WMAP map ($\hat{T}({\mathbf{x}})$).

Hence, these tests seem to indicate that systematic effects do not have a 
significant role in the non-Gaussian signature.

\subsection{Foregrounds}
\label{subsec:fore}
Foregrounds have been considered as the next possible source of non-Gaussianity.
We do not consider emission from point sources
because the brightest radio sources have been previously masked and, moreover, 
the angular scale of the non-Gaussian detection is much larger than the WMAP 
angular resolution. As regards the expected Sunyaev-Zel'dovich effect 
contribution due to galaxy clusters, it has been shown to be almost
negligible (Bennett et al. 2003b).

In order to study the frequency dependence of the non-Gaussian detection we 
have independently analysed the combined frequency maps:
$\hat{T}_Q({\mathbf{x}})$, $\hat{T}_V({\mathbf{x}})$ and 
$\hat{T}_W({\mathbf{x}})$.
For the frequency range of Q, V and W (from 41 GHz to 94 GHz), the non-Gaussian 
pattern found for the kurtosis does not resemble any of the frequency dependence 
due to the Galactic foregrounds (synchrotron, free-free and thermal dust).

\begin{figure}
\begin{center}

   \includegraphics[width=12cm]{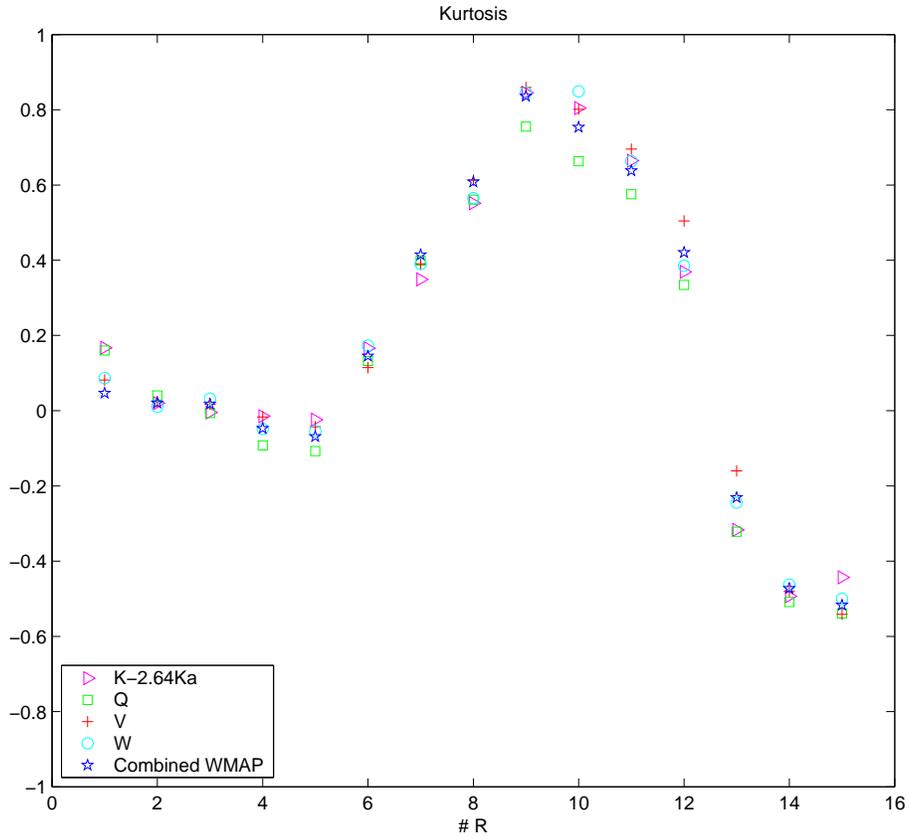}
 
\end{center}
\caption{\label{fig:fore1}Kurtosis values $K(R)$ obtained from
the analysis of the maps at each frequency maps. For reference, the kurtosis values
of the combined WMAP map are also included. 
It is clear that for the whole WMAP frequency range (from 23 GHz to 94 GHz) this
pattern is the same, indicating
that there is not frequency dependence for our non-Gaussian detection.
The kurtosis values obtained for the combination map $K - 2.65Ka$ (see text for
details) are also included.}
\end{figure}

We have also studied the clearly foreground contaminated K and Ka channels in 
this test. The pattern of the kurtosis curve for these channels does 
not completely follow the one for the other WMAP channels. Whereas the peak 
around $R_8$ clearly appears, the behaviour is very different at scales below 
$R_6$. In addition, there is an offset along all the scales. We think
that this  
effect corresponds to the very high foreground contamination that these 
channels show. In order to clarify this point, we have done the
following exercise. We  
have generated an additional CMB map by subtracting the Ka map (at 33 GHz) from 
the K one (at 23 GHz), multiplying the first one by a factor of 2.65. This 
number corresponds to the expected increment of the synchrotron emission from 33 
GHz to 23 GHz\footnote{A power law is assumed for the frequency dependence of 
the synchrotron emission: $T_{syn}(\nu) \propto T_{syn}(\nu_0) 
\Big(\nu/\nu_0\Big)^{-2.7}$, as it was proposed by Bennet et al. (2003b).}. 
As seen in Figure~\ref{fig:fore1}, the pattern of the kurtosis curve 
for this map is the same as the one detected for the combined WMAP map 
($\hat{T}({\mathbf{x}})$).
Therefore, the same pattern for the kurtosis is found, not only for the Q, V and 
W bands, but also for the whole WMAP frequency range (from 23 GHz to 94 GHz). 
This implies not only that the synchrotron does not seem to be the source of
the non-Gaussian signature, but also that this detection does not show any 
significant frequency dependence.

\begin{figure}
\begin{center}

   \includegraphics[width=12cm]{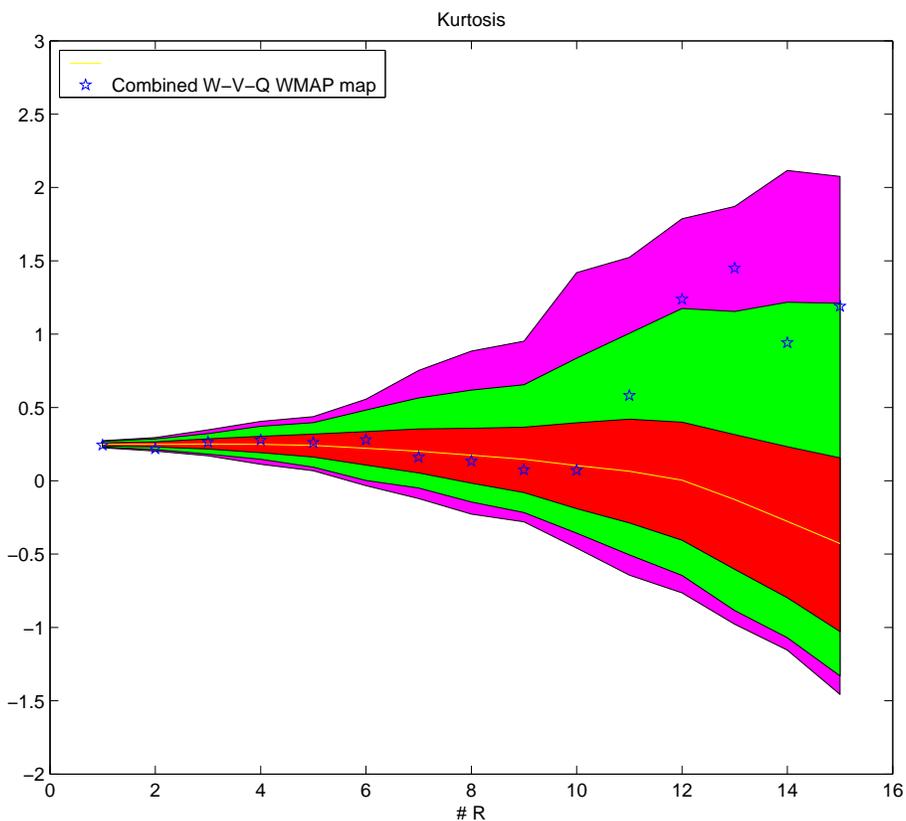}
 
\end{center}
\caption{\label{fig:fore2}Kurtosis values $K(R)$ obtained from a combined  
map defined by subtracting the Q and V receivers from the W ones. This map is almost
free
of CMB emission.
The acceptance intervals at the 32\% (red, inner), 5\% (green, middle) and
1\% (magenta, outer) significance levels
and the mean value given by the 1000 simulations (yellow solid line) for this
combined  
$W-V-Q$ map are also plotted. This CMB cleaned map seems to be compatible with the
Gaussian hypothesis, indicating that the remaining foreground contribution is not
high enough to be detectable trough our Gaussian test.}
\end{figure}

An additional test to check the influence of the foregrounds was done by 
analysing a map where the CMB contribution is small: the two receivers of the Q-
Band and the two ones of the V-Band are subtracted from the four receivers of 
the W-Band. In this map we have significant contributions from the foregrounds 
and the noise. The SMHW analysis of this map also shows (Figure~\ref{fig:fore2}) 
that, apparently, the foregrounds are not the source of the non-Gaussian 
detection since they are compatible with their own Gaussian bands.
Even more, the pattern of the kurtosis curve is completely different from the 
one estimated from the combined WMAP map ($\hat{T}({\mathbf{x}})$).

\begin{figure}
\begin{center}

   \includegraphics[width=12cm]{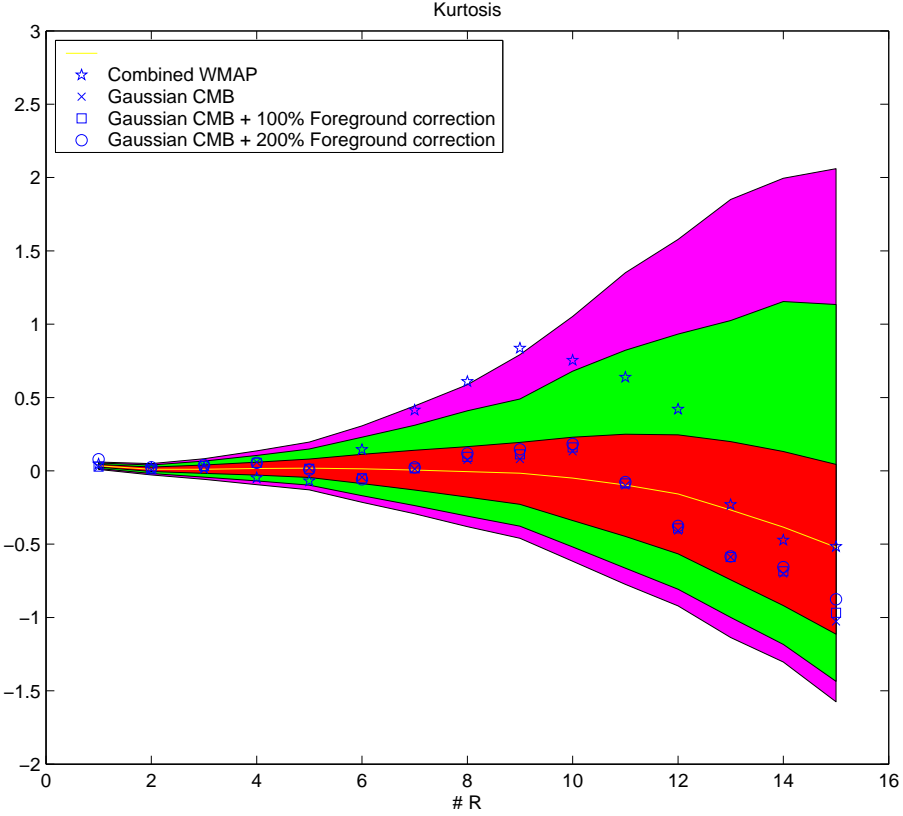}
 
\end{center}
\caption{\label{fig:fore3}The kurtosis values $K(R)$ obtained from 
artificially contaminated CMB Gaussian maps are presented, together with
the ones calculated for the combined WMAP (blue asterisks).
Two different degrees of contamination have been considered by adding to
a Gaussian simulation once and twice the foreground contribution obtained
by the WMAP team.
The acceptance intervals at the 32\% (red, inner), 5\% (green, middle) and
1\% (magenta, outer) significance levels 
and the mean value given by the 10000 simulations (yellow solid line) for the
combined  WMAP 
map ($\hat{T}({\mathbf{x}})$) are also given for a better comparison. }
\end{figure}

Finally, we have done the following exercise. 
We have estimated the foreground correction ($F({\mathbf{x}})$) subtracted from 
the data by the WMAP team just by doing:  $F({\mathbf{x}}) = T({\mathbf{x}}) - 
\hat{T}({\mathbf{x}})$.
This foreground correction is added to a Gaussian simulation of the combined 
WMAP map. The SMHW analysis was performed in this contaminated map. As shown in 
Figure~\ref{fig:fore3}, the curve for the kurtosis ($K(R)$) not only has a 
pattern completely different from the one detected in the combined WMAP map 
($\hat{T}({\mathbf{x}})$) but is also very similar to the uncontaminated Gaussian
simulation.
Even more, the foreground correction ($F({\mathbf{x}})$) was added 
twice to the Gaussian simulation, finding the same behaviour.
This exercise ---as the previous ones--- also seems to discard the Galactic 
foreground emissions as the ones responsible for the non-Gaussian detection.
\subsection{Influence of the uncertainties in the cosmological parameters}
\label{subsec:param}

Since we use CMB Gaussian simulations following the power spectrum
generated by  
CMBFast, we must also check how the uncertainties in the cosmological
parameters   
(and, hence, in the power spectrum) could affect the results. In
particular, how  
the 1\% significance bands depend on these uncertainties? 
An exhaustive analysis would require a very sophisticated statistical
framework that can accomodate uncertainties of cosmological parameter
estimates. 
Instead we have adopted two simplified approaches that, nevertheless,
we believe are good indicators of the effect on our results from the
uncertainties in the $C_\ell$'s.

\begin{figure}
\begin{center}

   \includegraphics[angle=270,width=16cm]{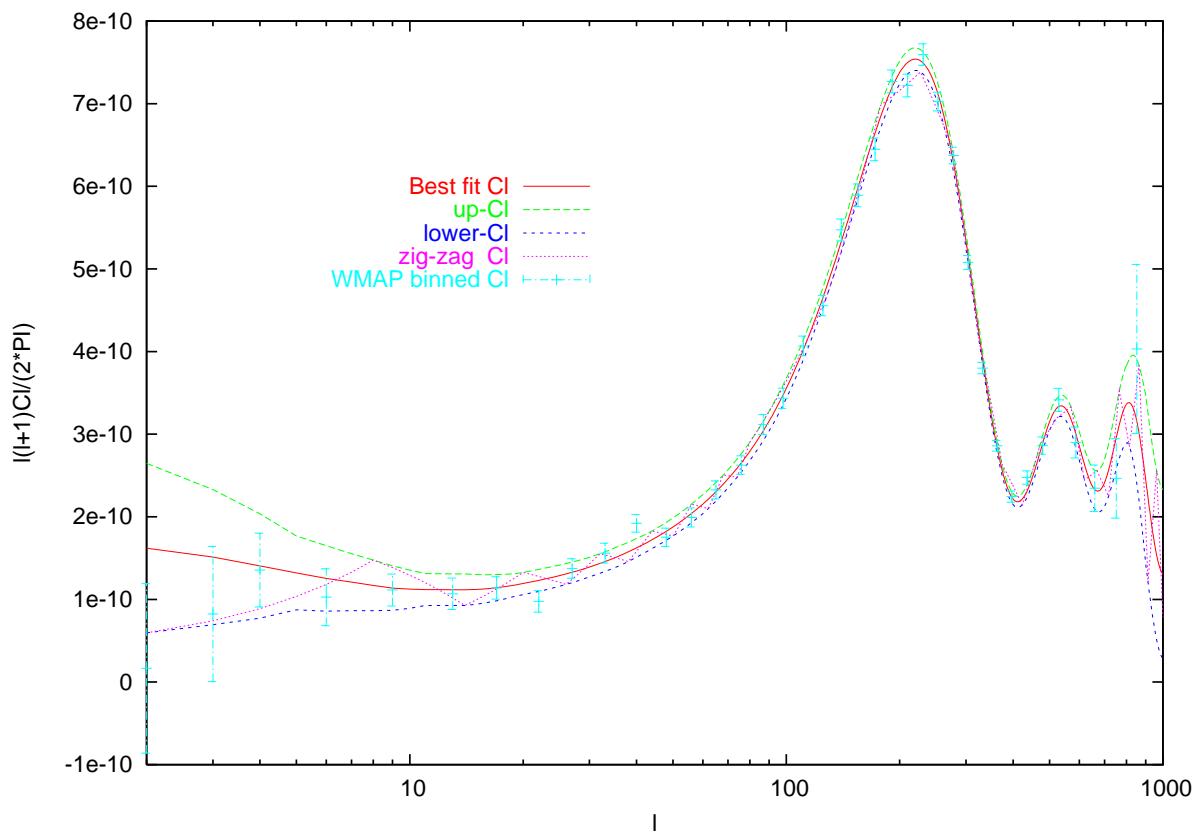}
 
\end{center}
\caption{\label{fig:cls1}Power spectra used to the test the
uncertainties in the 
cosmological
paramters assumed in this work. The bes-fit power spectrum given by the WMAP team is
plotted (solid red line)
together with the power spectrum estimated by the WMAP team (cyan points and error
bars). Three addional models are used:
the so-called 'upper-limit' and 'lower-limit' power spectra are constructed from the
best-fit model
by adding or subtracting $1\sigma$ and the so called 'zig-zag' model which oscillates
around the
best-fit model, always within the $1\sigma$ error.}
\end{figure}

\begin{figure}
\begin{center}

   \includegraphics[width=6cm]{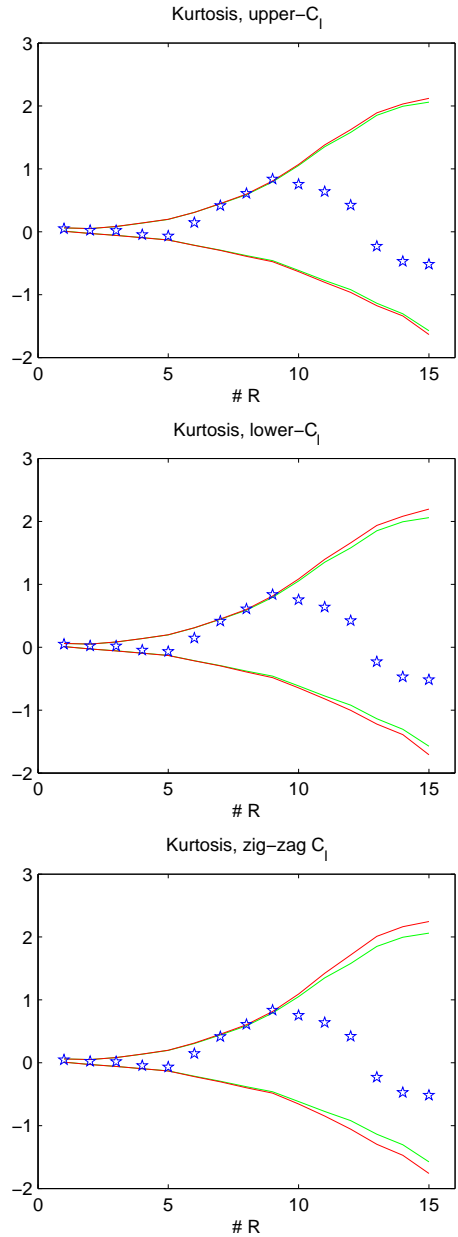}
 
\end{center}
\caption{\label{fig:cls2}Modification of the acceptance
interval of the kurtosis at the 1\% significance level,
due to the uncertainties in the power spectrum estimation. The 1\%
significance band obtained with the best-fit model
is plotted (green line) together with the one obtained with the
'upper-limit' power spectrum (upper panel,
red line),  the 'lower-limit' power spectrum (middle panel, red line) and the
'zig-zag' power spectrum
(lower panel, red line).}
\end{figure}

First, we
have performed Gaussian simulations with three different power
spectra, being  
all of them compatible with the $1\sigma$ errors of the WMAP estimated power 
spectrum. In Figure~\ref{fig:cls1} the power spectrum given by CMBFast using the 
best-fit Cosmological Model of the WMAP team (solid red line) and the power 
spectrum estimation for the WMAP (cyan) have been plotted. The three 
additional power spectra are defined as follows: the so-called 'upper-limit' and 
'lower-limit' power spectra are created from the best-fit power spectrum by 
adding (or subtracting) the $1\sigma$ error estimated by the WMAP team; the so-
called 'zig-zag' power spectrum is an intentionally exotic power spectrum that 
oscillates around the best-fit one, always within the 'upper-limit' and 'lower-
limit' power spectra.
By analysing Gaussian simulations following these power spectra, we can estimate 
the influence of the uncertainties of the power spectrum determination on the acceptance intervals at the 
$1\%$ significance level  
already established for the best-fit WMAP model. This is 
plotted in Figure~\ref{fig:cls2}. It can be seen that for small and intermediate 
scales (including the one at which the non-Gaussian detection is found), the 
uncertainties in the 1\% significance bands are negligible. Only at large scales we can find 
some deviations (especially for the ´zig-zag´ power spectrum).

The second approach to test the influence of the uncertainties in the 
cosmological parameters has been done by whitening the data as well as the 
simulations.
By doing this, most of the influence of the input power spectrum is eliminated. 
In other words, we also detect non-Gaussianity
wherever the original power 
spectrum was. In this case, we also obtain a non-Gaussian feature with
the same probability around the same scales.

\subsection{Is it due to Intrinsic fluctuations?}
\label{subsec:intrinsic}
The previous Subsections have shown that there are not strong evidences for the 
non-Gaussian detection due to neither systematic effects nor Galactic foreground 
emissions nor uncertainties in the cosmological parameters.
Therefore, other sources for this detection may be considered. 

\begin{figure*}
\begin{center}

   \includegraphics[angle = 270, width=8cm]{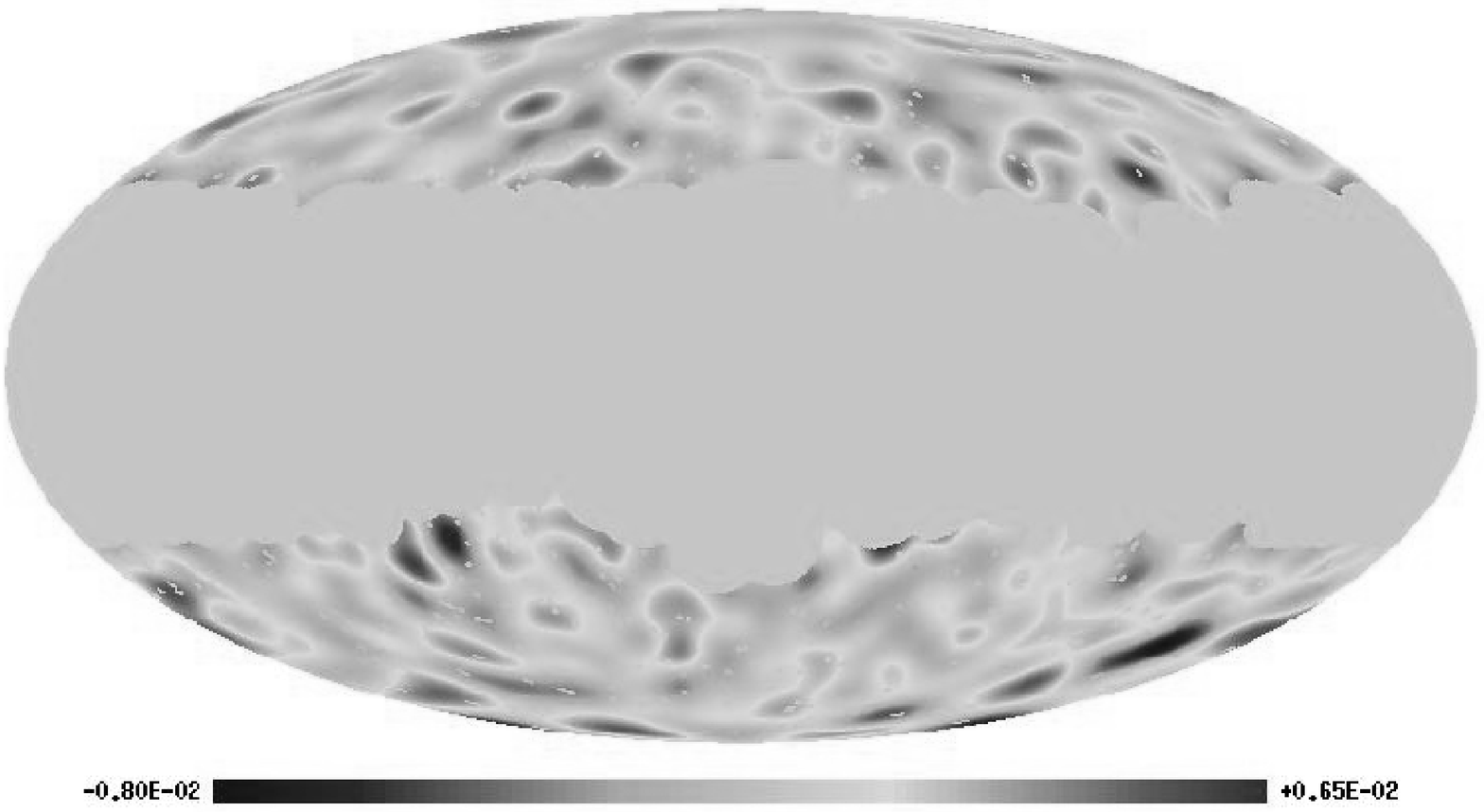}
   \includegraphics[angle = 270, width=8cm]{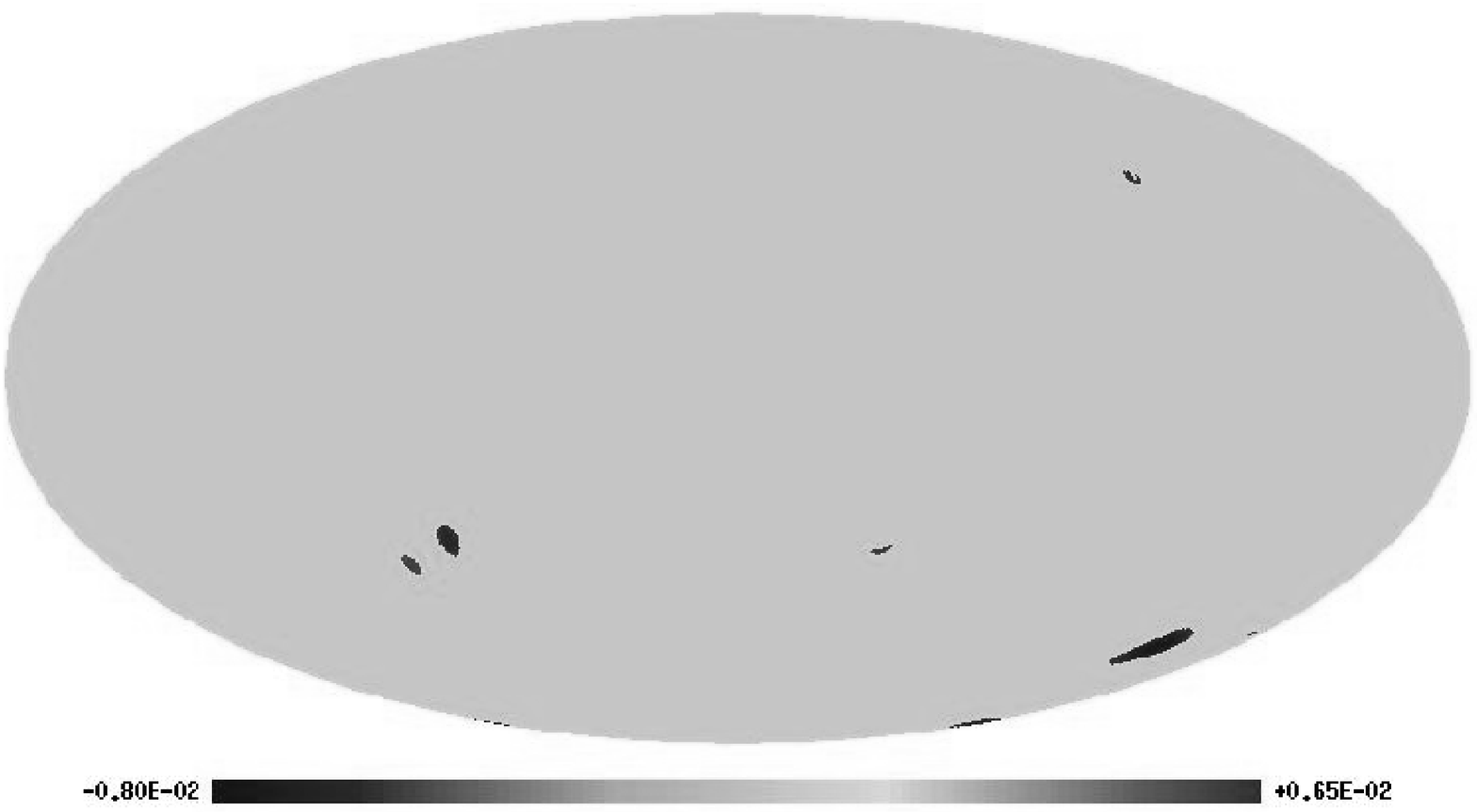}
 
\end{center}
\caption{\label{fig:sky}In the left panel, the SMHW coefficients at $R_8 = 250.0$
arcmin outside
the \emph{exclusion mask} $M(R_8)$ are presented. In the right panel, only those
coefficients above $3\sigma(R_8)$
are plotted. If these coefficients are not considered, the kurtosis of the remaining
ones is completely
compatible with the Gaussian model. The coldest (blue) spot at $b = -57^\circ, l =
209^\circ$ has a minimum
value equals to $-4.57\sigma(R_8)$. The simulations indicate that the probability of
this value is $\approx 1 \%$.}
\end{figure*}

\begin{figure*}
\begin{center}

   \includegraphics[width=16cm]{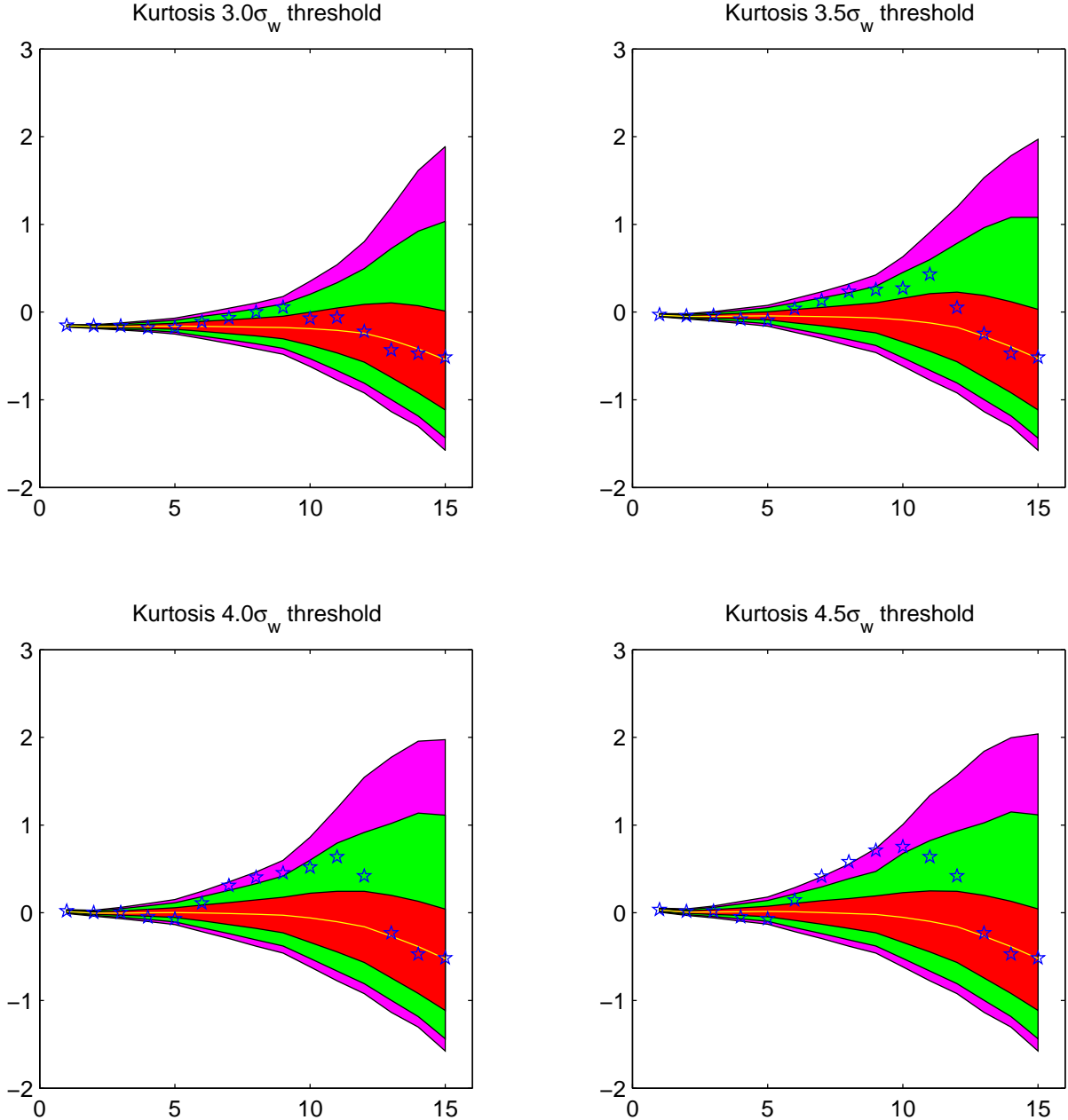}

\end{center}
\caption{\label{fig:sigmas}Kurtosis values for different thresholds of
the wavelet coefficients. From left to right and top to bottom, only
thoses pixels below
$3\sigma_w$, $3.5\sigma_w$, $4\sigma_w$ and $4.5\sigma_w$ are considered. This
process is done
for both data (blue stars) and simulations. The 
acceptance intervals
at the 32\% (red, inner), 5\% (green, middle) and 1\% (magenta,
outer) significance levels
and the mean value given by the 10000 simulations (yellow solid line) for the
combined  WMAP 
map ($\hat{T}({\mathbf{x}})$) are also given for a better comparison. As the
coefficients with the most extreme
values are not considered, the WMAP data get more compatible with the Gaussian
hypothesis.}

\end{figure*}

In particular, intrinsic fluctuations can not be rejected.
One of the powerful advantages of the wavelet analysis is the possible 
identification, on the sky, of the non-Gaussian source. In Figure~\ref{fig:sky} 
(left panel) the SMHW coefficients at the scale $R_8$ are plotted. 
The minimum value of this map is $-4.57\sigma(R_8)$, where $\sigma(R_8)$ is the 
dispersion of the wavelet  coefficients at scale $R_8$. Using the 10000 Gaussian 
simulations, we have checked that the probability for an extreme like this is
$\approx 1\%$. The right panel of Figure~\ref{fig:sky} shows those 
SMHW coefficients that are above (in absolute value) $3\sigma(R_8)$.
If these pixels are not taken into account in the kurtosis calculation, the 
estimation for this statistic is perfectly compatible with the Gaussian 
simulations.
In fact, we have computed the acceptance intervals at the 32\%, 5\% 
and 1\% significance levels for 
additional sets of 10000 Gaussian simulations, where those pixels with values 
above a given threshold are not taken into account in order to compute the 
kurtosis curve $K(R)$.
The thresholds are: $3.0\sigma_w(R)$, $3.5\sigma_w(R)$, $4.0\sigma_w(R)$ and 
$4.5\sigma_w(R)$.
These acceptance intervals have been used to check if the combined WMAP map 
satisfies the Gaussian hypothesis for each one of these different
thresholds. The  
results are presented in Figure~\ref{fig:sigmas}. It is clear that, as the 
threshold decreases, the combined WMAP data is more compatible with Gaussianity. 
In particular, for the $3\sigma_w$ threshold the WMAP data is compatible with 
the Gaussian hypothesis (within the acceptance interval at the 5\% 
significance level).

\begin{figure}
\begin{center}

   \includegraphics[angle=270, width=15cm]{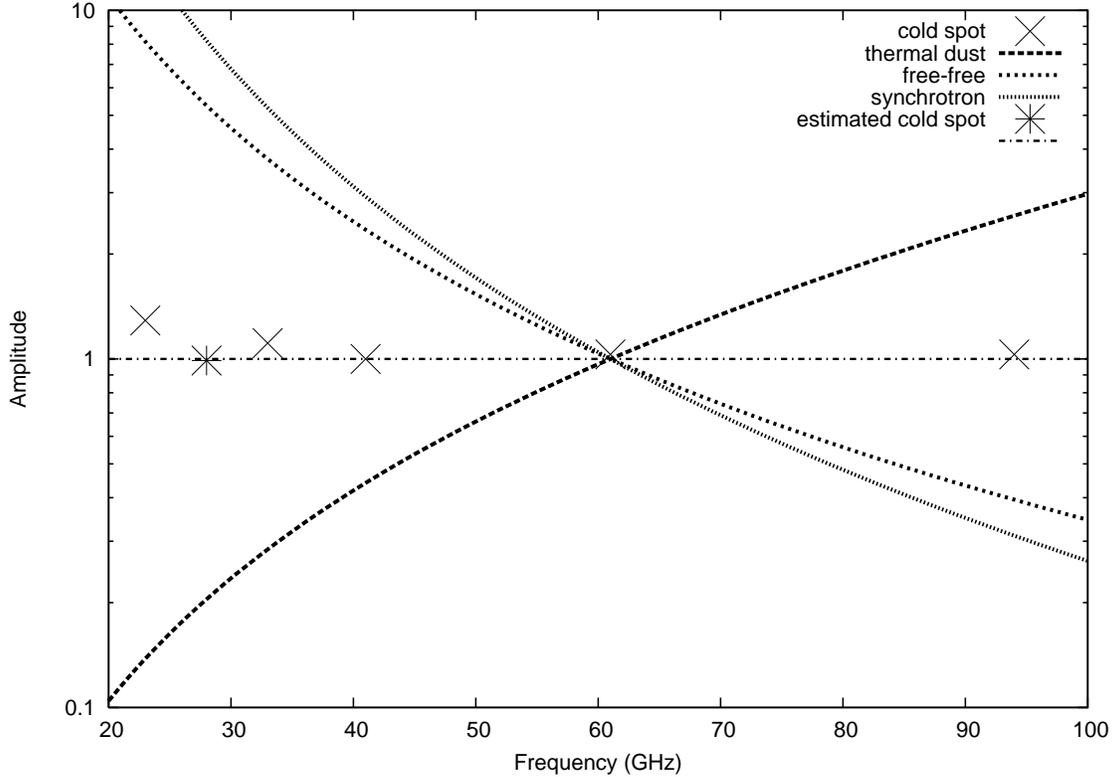}
 
\end{center}
\caption{\label{fig:freq}The mean frequency dependence followed by the
300 most negative SMHW coefficients in the cold spot
at $b = -57^\circ, l = 209^\circ$ (crosses) is plotted together 
with the expected frequency dependence for the
different foregrounds, at Galactic latitudes $-90^\circ < b < -57^\circ$:
synchrotron, free-free and thermal dust.
The asterisk represents the cold spot, after the Ka map
(corrected by the synchrotron factor, as indicated in the text) 
is subtracted from the K one. 
The dot-dashed line has been plotted for comparison with a perfect CMB
behaviour. All the amplitudes are normalised to one at
the Q-band. Notice that  
the frequency dependence
for the different foregrounds
is preserved in wavelet space,
due to the linearity of the wavelet
convolution.}
\end{figure}

A very interesting exercise was done in order to study the previous very cold 
spot (in SMHW space) in the southern hemisphere ($b = -57^\circ, l = 
209^\circ$). We have calculated the mean value of the cold spot (at $R_8$)
at each one of the WMAP frequency channels. As shown in 
Figure~\ref{fig:freq}, the frequency dependence of this spot follows the CMB 
one, and is not compatible with any of the Galactic foregrounds. Notice that if 
the 2.65 times Ka map is subtracted from the K one, we get an even better 
agreement with constant frequency dependence, given by the dot-dashed line. We 
have also studied the frequency dependence of this cold spot in real space 
(after filtering all the frequency maps with a Gaussian beam of FWHM = 
$8^\circ$, the size corresponding to the SMHW scale $R_8$). A good agreement 
with the CMB behaviour is found for bands Q, V and W, whereas a clear 
contamination due to synchrotron appears between bands K and Ka. However, when 
the 2.65 times Ka map is subtracted from the K one, the CMB behaviour is 
recovered.
We notice that whereas for the K and Ka bands in real space  the synchrotron 
emission dominates over CMB, this is not the case for K and Ka bands in wavelet 
space, since the SMHW has diminished the synchrotron amplitude relative to the
CMB one (due to the large scale variation of the former).
Hence, the frequency analysis performed on the cold spot, also indicates that 
the foregrounds do not seem to be the source of the non-Gaussianity detection.
This cold spot appears as a real structure in the sky, not only seen by WMAP, 
but also by COBE-DMR. If the COBE-DMR map is convolved with the SMHW at $R_8$, 
we find some pixels above $3\sigma_w$ around this point.

There are several possibilities which can explain the non-Gaussian detection for 
the kurtosis of the SMHW coefficients.
For instance, massive superstructures (like superclusters or Great
Attractor-like structures) or large voids, can produce secondary anisotropies 
through the Rees-Sciama effect (Mart{\'\i}nez-Gonz{\'a}lez \& Sanz 1990).
Even more, also primordial anisotropies can be considered. Non standard 
inflationary models could also explain the detected non-Gaussianity (see e.g. 
Acquaviva et al. 2002 and references therein). In addition,  some kind of 
topological defects  like monopoles or textures (Turok \& Spergel 1990) could be 
present in the sky. On the contrary, non isotropic topological defects like 
cosmic strings may not produce this kind of non-Gaussianity, since their 
characteristic scale is around arcminutes and also the detection has been done 
with an isotropic wavelet.

\section{Conclusions}
\label{sec:final}
We have presented the detection of non-Gaussianity in the WMAP 1-year
data, in  
the kurtosis of the SMHW coefficients at scales around $4^{\circ}$,
which  
implies a size in the sky of around $10^{\circ}$. 
At those scales, the kurtosis values corresponding to the WMAP
combined 
map are outside the acceptance interval at the 1\% significance level,
with a right tail probability of $\approx 0.4\%$.
The Gaussianity study was also done on both hemispheres 
independently, showing that the northern hemisphere is compatible with
the  
Gaussian model, whereas the southern one presents a very clear
non-Gaussian  
signal similar to the one detected in the all-sky analysis. This
detection has a right tail probability of $\approx 0.1\%$
The number of performed simulations has been checked to be enough to
establish accurately the acceptance interval at the 1\% significance 
level for the kurtosis
values at the different SMHW scales studied.

The independent analyses performed on each hemisphere 
also show that the northern hemisphere seems to have 
less structure than the southern one at scales between $4^\circ$ and $12^\circ$, 
whereas the southern one has less structure than the northern hemisphere at 
larger scales. However, both scalograms, the one for the north and the one for 
the south, seem to be compatible with the Gaussian model. Our non-Gaussian 
detection differs from other asymmetries reported up to date (Park 2003 and 
Eriksen et al. 2003), since a direct non-Gaussian signature has been found in 
the kurtosis of the wavelet coefficients at scale $R \approx 10^\circ$.

Since those SMHW pixels clearly affected by the mask decrease the efficiency 
of the analysis, only those pixels with a small contribution from the
Galactic mask (\emph{Kp0}), at each SMHW scale, have been analysed.
However, the non-Gaussian detection is quite insensitive to the particular 
choice of the \emph{exclusion masks}, since we have tested that it appears outside
the acceptance intervals at 
the same significance level for several choices.

We have performed several tests in order to identify the source of
this non-Gaussian signature. Systematic effects and foregrounds have
been carefully 
studied. The different channels have been checked, showing that all the CMB 
dominated ones (Q, V and W) have the same pattern for the kurtosis of
the SMHW coefficients.
Moreover, by analysing the maps produced by subtracting the data given by the 
receivers at each frequency ---these maps are almost free of CMB and
foreground  
contributions--- we have shown that the non-Gaussian signal disappears.
The non-Gaussian signal appears in the whole WMAP frequency range
(from 23 GHz  
to 94 GHz) showing no frequency dependence. 
Even more, we added overestimated foreground contamination to Gaussian
CMB maps,  
showing that the detected non-Gaussian signal does not appear in the SMHW 
analysis.
This seems to indicate no correlation with Galactic foregrounds. Since the 
brightest point sources are masked and the non-Gaussian features appear at 
intermediate scales, we do not expect a significant contribution coming from 
this emission. The Sunyaev-Zel'dovich effect due to galaxy clusters is 
negligible for the WMAP 1-year data (only the most prominent local
clusters like Coma can contribute at the $2\sigma$ level).

The uncertainties in the power spectrum estimation (and, hence, in the 
cosmological parameters) have also been taken into account, showing that the 
1\% significance bands for the kurtosis (at the scales of interest) are not affected by such  
uncertainties. Three additional power spectra covering the $1\sigma$
error band  
in the power spectrum determination were used.
Even more, an additional test was done by performing a whitening of the maps.
The non-Gaussian signal is also detected in this case.
This gives robustness to the result, since possible errors in the
estimation of the $C_\ell$ of the Gaussian model do not affect it.

Thanks to the properties of the SMHW, we can identify (in the sky) the possible 
source of the non-Gaussianity. We have found a cold spot in the southern 
hemisphere (in $b = -57^\circ, l = 209^\circ$) with a probability $\approx 1\%$. 
The frequency dependence of this spot is compatible with the one of
the CMB. A more detailed study of extrema will be presented in a 
future work. This study should be done in both real and wavelet 
spaces and could be very useful in order to clarify if the non-Gaussian 
signature is localized in certain positions on the sky or otherwise, it is 
produced by a non-Gaussian temperature distribution.

Finally, taking into account all the performed tests, intrinsic fluctuations 
(due to secondary anisotropies like the Rees-Sciama effect or due to primordial 
fluctuations produced by topological defects or non-standard inflation) cannot 
be rejected. The WMAP 2-year data will be, undoubtedly, very useful to confirm 
or not these results.

\acknowledgments
Authors kindly thank Dr. Eiichiro Komatsu for very useful comments regarding the 
data reduction and suggestions and Julio Edgar Gallegos for useful discussions.
PV thanks Universidad de Cantabria (UC) for a postdoc grant.
RBB thanks UC and Spanish Ministerio de Ciencia y Tecnolog{\'\i}a (MCYT) for a 
Ram{\'o}n y Cajal contract.
We acknowledge partial financial support from the Spanish MCYT project ESP2002-
04141-C03-01.
We kindly thank IFCA for providing us with its Grid Wall cluster to generate and 
analyse the WMAP simulations. We also thank Centro de Supercomputaci{\'o}n de 
Galicia (CESGA) for providing the Compaq HPC320 supercomputer to run part of the 
analysis. 
We acknowledge the use of the Legacy Archive for Microwave Background Data 
Analysis (LAMBDA). Support for LAMBDA is provided by the NASA Office of Space 
Science.
This work has used the software package HEALPix (Hierarchical, Equal
Area and iso-latitude pixelization of the sphere,
http://www.eso.org/science/healpix), developed by K.M. Gorski,
E. F. Hivon, B. D. Wandelt, J. Banday, F. K. Hansen and
M. Barthelmann.
We acknowledge the use of the software package CMBFAST
(http://www.cmbfast.org) developed by U. Seljak and M. Zaldarriaga.

\clearpage

\clearpage

\clearpage

\clearpage

\clearpage

\clearpage

\clearpage

\clearpage

\clearpage

\clearpage

\clearpage

\clearpage

\clearpage

\end{document}